\documentclass[useAMS,usenatbib]{mn2e}
\usepackage{graphicx}
\usepackage{amsmath}
\usepackage{breqn}
\usepackage{color}
\usepackage{amssymb}
\usepackage{epsfig}
\usepackage{natbib}
\usepackage{bmpsize}
\title[Unabsorbed type-2 AGN]{AGN with discordant optical and X-ray classification are not a physical family: Diverse origin in two AGN}
\author[I. Ordov\'as-Pascual ]{
I. Ordov\'as-Pascual$^{1}$\thanks{E-mail:
ordovas@ifca.unican.es},
S. Mateos$^{1}$,
F. J. Carrera$^{1}$,
K. Wiersema$^{2}$,
X. Barcons$^{1}$,
\newauthor V. Braito$^{3}$,
A. Caccianiga$^{4}$,
A. Del Moro$^{5}$,
R. Della Ceca$^{4}$,
P. Severgnini$^{4}$\\
$^{1}$Instituto de F\'isica de Cantabria (CSIC-UC), E-39005, Santander, Spain\\
$^{2}$Department of Physics and Astronomy, University of Leicester, Leicester, UK, LE1 7RH\\
$^{3}$INAF - Osservatorio Astronomico di Brera, via E. Bianchi 46, I-23807 Merate, Italy\\
$^{4}$INAF - Osservatorio Astronomico di Brera, Via Brera 28, I-20121, Milan, Italy\\
$^{5}$Max-Planck-Institut f\"{u}r Extraterrestrische Physik (MPE), Postfach 1312, D-85741 Garching, Germany}
\begin{document}
\date{Accepted \dots . Received \dots ; in original form \dots}
\pagerange{\pageref{firstpage}--\pageref{lastpage}} 
\maketitle

\label{firstpage}
\begin{abstract}
Approximately 3-17 percent of Active Galactic Nuclei (AGN) without detected rest-frame UV/optical broad emission lines (type-2 AGN) do not show absorption in X-rays. The physical origin behind the apparently discordant optical/X-ray properties is not fully understood. Our study aims at providing insight into this issue by conducting a detailed analysis of the nuclear dust extinction and X-ray absorption properties of two AGN with low X-ray absorption and with high optical extinction, for which a rich set of high quality spectroscopic data is available from XMM-Newton archive data in X-rays and XSHOOTER proprietary data at UV-to-NIR wavelengths. In order to unveil the apparent mismatch, we have determined the A$_{\rm V}$/N$_{\rm H}$ and both the Super Massive Black Hole (SMBH) and the host galaxy masses. We find that the mismatch is caused in one case by an abnormally high dust-to-gas ratio that makes the UV/optical emission to appear more obscured than in the X-rays. For the other object we find that the dust-to-gas ratio is similar to the Galactic one but the AGN is hosted by a very massive galaxy so that the broad emission lines and the nuclear continuum are swamped by the star-light and difficult to detect.

\end{abstract}
\section{Introduction}
\label{sec_intro}

The standard unified model of AGN (\citealt{antonucci93}, \citealt{urry95}) explains the observed differences between optical type-1 and type-2 AGN through orientation effects. If our line-of-sight to the central engine intercepts the nuclear absorber invoked by unified models, the UV/optical continuum emission, the rest-frame UV/optical broad emission lines (line widths $\geq$1500 km/s in Full Width at Half Maximum, FWHM) and the X-ray emission, originated at sub parsec scales, should be absorbed. In this case, the AGN is classified as type-2. On the contrary, if we have a direct view of the central engine, UV/optical broad emission lines should be detected, while the X-ray emission should have low absorption (N$_{\rm H}<$4$\times$10$^{21}$ cm$^{-2}$, the equivalent to A$_{\rm V}$=2 mag using a Galactic dust-to-gas ratio, \citealt{caccianiga08}). In this case, the AGN is classified as type-1.

The classification of AGN using either the optical range or X-rays should agree according to this model. Nevertheless, approximately 10-23 percent of AGN optically classified as type-1 present an X-ray absorbed spectrum (normally with N$_{\rm H}<10^{22}\, {\rm cm}^{-2}$), while 3-17 percent of type-2 AGN are X-ray unabsorbed (eg. \citealt{panessa02}, \citealt{caccianiga04}, \citealt{mateos05a}, \citeyear{mateos05b}, \citealt{mainieri05}, \citealt{caccianiga08}, \citealt{mateos10}, \citealt{corral11}, \citealt{scott12}, \citealt{page12}, \citealt{merloni14}). The mismatch between optical extinction and X-ray absorption described above is observed in both optical/infrared and X-ray selected samples at all redshifts. The origin of such apparent discrepancies remain unclear, as well as the validity of the unified model for such AGN. To unveil the nature of such discrepancies, we need detailed studies on these discordant AGN.

For X-ray unabsorbed type-2 AGN, one possibility to explain this discrepancy can be the presence of a Compton-thick absorber (intrinsic  N$_{\rm H}$ equal or larger that the inverse of the Thomson cross-section: N$_{\rm H}>\sigma_{\rm T}^{-1}$=1.5$\times$10$^{24}$ cm$^{-2}$). In this case the direct X-ray emission below 10 keV should be completely suppressed and we would only detect scattered nuclear radiation (\citealt{braito03}, \citealt{akylas09}, \citealt{braito11}, \citealt{malizia12}). Since the scattered emission is only 1-3 per cent of the intrinsic AGN emission (\citealt{gilli01}, \citealt{comastri04}, \citealt{georgantopoulos11a}), the sources would be identified as low luminosity, unabsorbed type-2 AGN. Another possibility is that the broad UV/optical lines are diluted by the host galaxy emission (\citealt{severgnini03}, \citealt{georgantopoulos05}, \citealt{caccianiga07}, \citeyear{caccianiga08}). An alternative explanation could be a high dust-to-gas ratio: normally, AGN  show dust-to-gas ratios below the Galactic standard or comparable (\citealt{maiolino01}, \citealt{vasudevan09}, \citealt{parisi11}, \citealt{marchese12}, \citealt{hao13}, \citealt{burtscher16}), if this ratio is substantially higher, significant suppression of the broad line emission could take place without strong effects in the X-ray bands. Dust-to-gas ratios well above the Galactic value have been found in some AGN, although such cases are rare (\citealt{caccianiga04}, \citealt{trippe10}, \citealt{huang12}, \citealt{malizia12}, \citealt{masetti12}, \citealt{mehdipour12}). In the sample of \cite{maiolino01}, a sample of AGNs whose X-ray spectrum shows cold absorption and whose optical and/or IR spectrum show at least two broad lines, this is found in 9 per cent of the sources, and in the HBS28 sample (\citealt{caccianiga04}) this is only found in 3 per cent. In other objects optical observations show an intrinsically high Balmer decrement for the Hydrogen broad emission lines, while the X-ray spectra show low absorption \citep{barcons03}. A dusty-ionized absorber like the one in NGC 7679 \citep{della01} can produce more relative absorption in the X-rays than in the optical emission. One last possible explanation is a variability scenario since optical and X-ray observations are normally obtained at different epochs. However, we note that, even with simultaneous observations, there are some objects whose optical and X-ray classification do not match (\citealt{corral05}, \citealt{bianchi08}, \citeyear{bianchi12}).

The objective of this study is to get insight into the physics behind the apparent mismatch between UV/optical and X-ray classification of two low-z AGN selected
from the Bright Ultra-hard XMM-Newton Survey (BUXS, \citealt{mateos12}):
2XMMiJ000441.2+000711 (hereafter J00, z=0.1075, \citealt{abazajian09}) and
2XMMJ025218.5-011746 (hereafter J02, z=0.0246, \citealt{jones09}).
While both sources appear unabsorbed in X-rays,
J00 has been optically  classified as
type-1.9 using the SDSS-DR7 spectrum \citep{abazajian09} and J02 is a Sb edge-on galaxy \citep{vaucouleurs91} optically classified as a type-2 AGN (6dF spectrum, \citealt{jones09}).

For these sources, we obtained UV-to-NIR XSHOOTER spectra (P.I.: S. Mateos) and by comparing the properties derived in X-rays and in the UV/optical, we have tested three possible scenarios to explain the discordance: a) the presence of a Compton-thick AGN; b) these sources are a normal AGN but in a very massive host galaxy, or weak AGN in a normal host galaxy and their broad UV/optical emission lines are diluted by the host galaxy emission; c) intrinsic non-standard nuclear properties, such as a high dust-to-gas ratio or an intrinsically weak Broad Line Region (BLR).

This paper is organized as follows. In Section~\ref{sec_sample} we explain how our objects are selected. In Section~\ref{sec_x} we describe the XMM-Newton data and in Section~\ref{sec_obs} we describe the XSHOOTER observations. In Section~\ref{sec_analysis} we derive the AGN emission by subtracting the host galaxy emission, as well as the AGN intrinsic reddening, its emission lines and the host galaxy properties. We finally discuss all the possible contributions that can cause a mismatch between the X-ray and UV/optical properties of our objects in Section~\ref{sec_discuss}. Throughout this paper errors are 1$\sigma$. We assume a $\Lambda$CDM cosmology with $\Omega_{\rm M}$=0.3, $\Omega_\Lambda$=0.7 and H$_0$=70 km s$^{-1}$ Mpc$^{-1}$.

\section[]{The sample}
\label{sec_sample}

The two AGN analyzed in this work were selected from the wide-angle (44.43 sq. degree) Bright Ultra-hard XMM-Newton Survey \citep{mateos12}. This is a flux-limited sample of 255 AGN detected in the 4.5-10 keV band with the XMM-Newton observatory. The objects have relatively bright X-ray fluxes $f_{\rm 4.5-10\,keV}\geq$ 6$\times 10^{-14}$ erg s$^{-1}$ cm$^{-2}$. At the time of writing, the optical spectroscopic identification completeness is $>$98 percent. There are 111 AGN in BUXS with optical spectroscopic classifications 1.8, 1.9 or 2. Of these, the 9 objects showing low X-ray absorption (N$_{\rm H}<$4$\times$10$^{21}$ cm$^{-2}$; \citealt{caccianiga08}) that are visible from Paranal were included in a proposal for followup with XSHOOTER. Only the two objects with the lowest declination were successfully observed with XSHOOTER. We discuss here in detail the properties of these two objects.

For both sources, we have proprietary high resolution UV-to-NIR XSHOOTER spectra, as well as good  quality XMM-Newton X-ray spectra \citep{jansen01} in the observed energy range from 0.25 to 10 keV (Table~\ref{data} and Fig.~\ref{j0x}).

\section[]{X-ray properties}
\label{sec_x}

\begin{table*}
\tabcolsep=0.11cm
\caption{X-ray information about the selected objects.}
\begin{tabular}{ | c || c | c | c | c | c | c | c | c | c | c | c | c |}
\hline
Object &  z & Obs. ID & T. exp  & Cts & Flux  & log(L) & N$_{\rm H}$ & N$_{\rm H,G}$  & $\Gamma$ & kT & Model & $\chi^2$/d.o.f\\
(1) &  (2) & (3) & (4) & (5) & (6) & (7) & (8) & (9) & (10) & (11) & (12) & (13)  \\
\hline                    
 &  & 0305751001 & 26574 & 1172  &  &  & & &  &  & & \\
 J00 & 0.1075 & (2005-12-10) & (11476)   & (1318) & 2.04$\pm$0.22 & 42.76$\pm$0.05 & $<$0.67 & 0.31 & 1.66$\pm$0.09 & 0.16$^{+0.01}_{-0.02}$ & \textit{bb+po} & 169.2/139 \\
&  & 0151490101 & 55203 & 830 & & & & & & &  & \\
J02 & 0.0246 & (2003-07-16) & (22017) & (866)  & 1.31$^{+1.07}_{-0.99}$& 41.25$\pm$0.03 & 1.7$_{-1.4}^{+2.0}$  & 0.51 & 2.08$\pm$0.09 & - & \textit{apo} & 135.1/100 \\
\hline
\end{tabular}

$Notes$: (1): J00=2XMMiJ000441.2+000711, J02=2XMMJ025218.5-011746. X-ray source identifier as listed in the Second XMM-Newton Serendipitous
Source Catalogue (2XMM-DR3; \citealt{watson09}; http://xmmssc-www.star.le.ac.uk/Catalogue/xcat\underline{ }public\underline{ }2XMMi-DR3.html). (2): Redshift. (3): XMM-Newton Observation ID. In brackets we show the date of the observation. (4): Observation exposure time in seconds after removal of high background flares in MOS1+MOS2 and in pn in brackets. (5): Net counts of the MOS spectra between 0.25 and 10 keV and pn spectra between 0.25 and 10 keV, the latter in brackets. (6): 2-10 keV flux in units of 10$^{-13}$ erg cm$^{-2}$ s$^{-1}$. (7): Logarithm of luminosity in the 2-10 keV range corrected for extinction. (8): Best-fit X-ray column density in units of 10$^{21}$ cm$^{-2}$.; (9): Galactic column density from the Dickey \&~Lockman HI map \citep{dickey1990} in units of 10$^{21}$ cm$^{-2}$. (10): Power law photon index. (11): Temperature of the black body. (12): Best-fitting model, where \textit{bb+po} stands for black body emission plus a power law, and \textit{apo} stands for an absorbed power law. All reported errors are at the 1$\sigma$ level.
\label{data}

\end{table*}

 \begin{figure*}
 \centering
 \includegraphics[width=8.5cm]{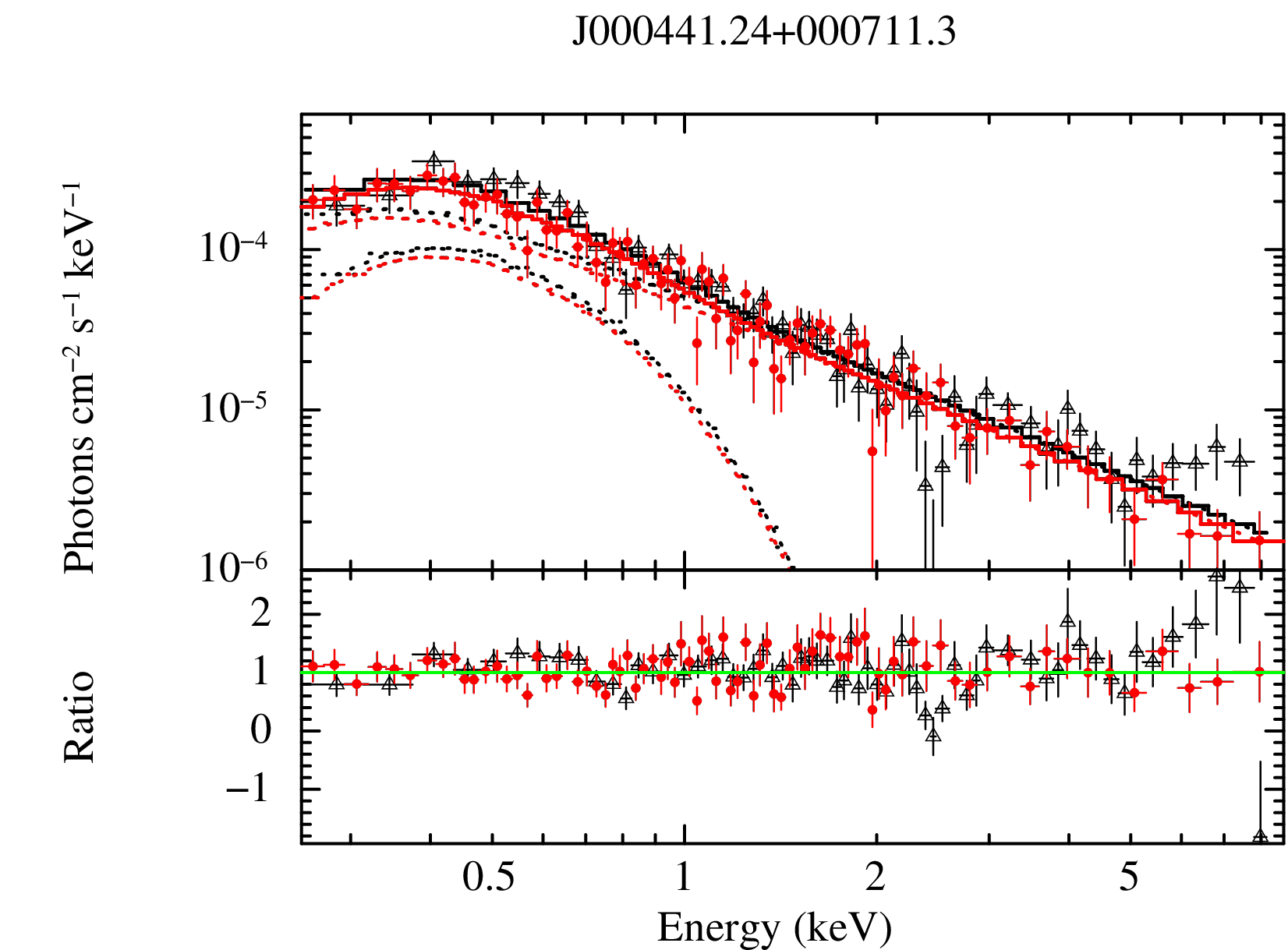}
 \includegraphics[width=8.5cm]{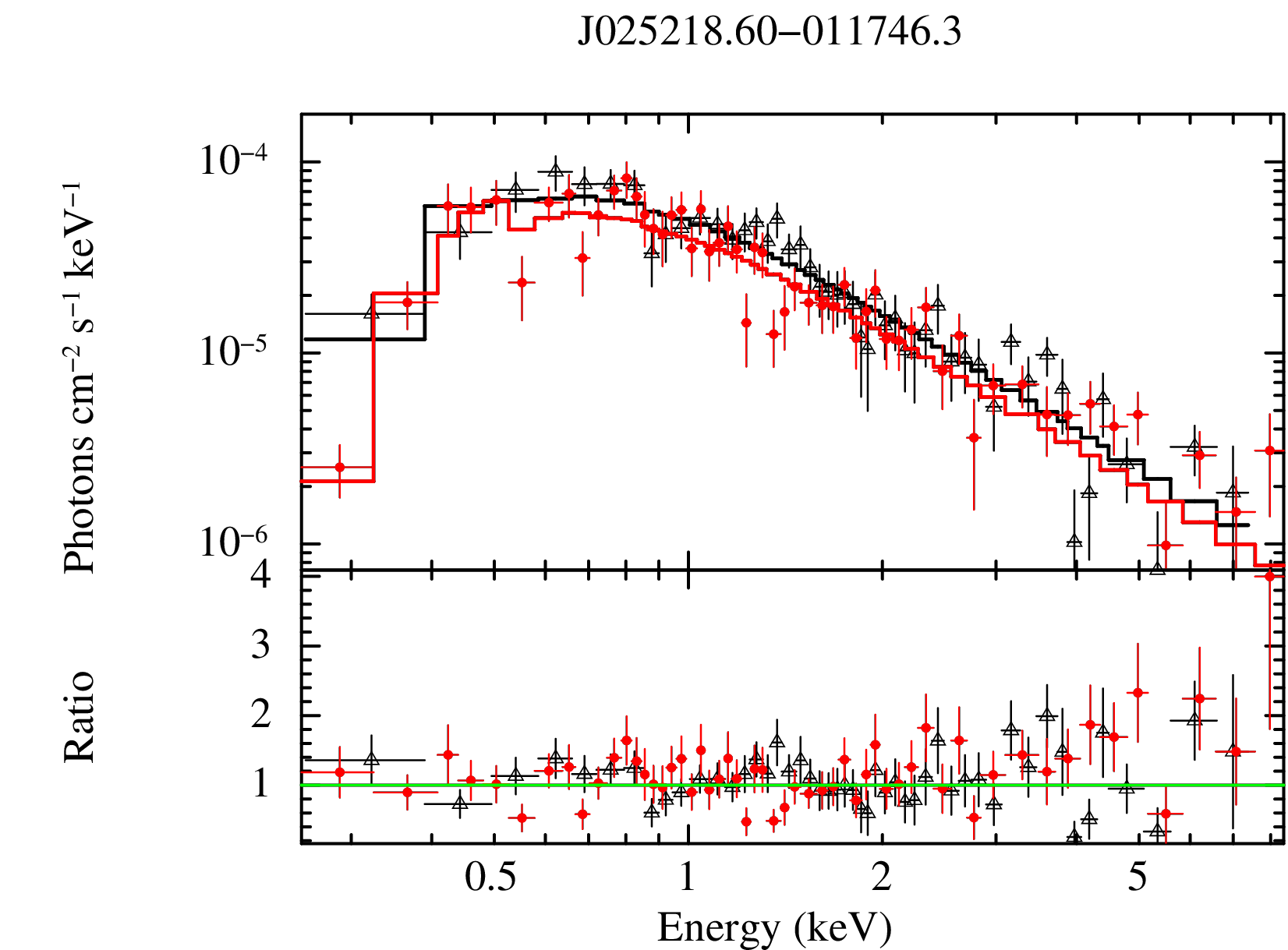}
    \caption{Left: MOS (black triangles) and pn (red dots) spectra of J00 and the best fit model (black-body + power law) in solid lines. Each component of the X-ray model is plotted with dotted lines. Right: MOS (black) and pn (red) X-ray spectra of J02 and the best fit model (absorbed power law) in solid lines. We also represent the ratio between the data and the best-fitting model. }
 \label{j0x}
 \end{figure*}

Source and background spectra were extracted at energies from 0.25 to 10 keV using circular regions. We used the XMM-Newton Science Analysis System (SAS) task \texttt{eregionanalyze} to obtain the circles that maximised the signal-to-noise ratio. For J00 we used radii of 34 and 32 arcsec for the MOS \citep{struder01} and pn \citep{turner01} cameras, respectively. For J02 we used a radius of 29 arcsec for MOS and 25 arcsec for pn. The background spectra were extracted
using circular regions of 50 arcsec radius located
in source free regions in the same CCD chip as
our objects. The response matrices and effective
area curves were obtained using the SAS tasks \texttt{rmfgen} and \texttt{arfgen}, respectively. We combined MOS1 and MOS2 source and background spectra and the corresponding response matrices.
The spectra were grouped with a minimum of 15 counts per bin and are shown in Fig.~\ref{j0x}.

The X-ray spectroscopic analysis was conducted with the \texttt{XSPEC} \textnormal package (v12.9.1; \citealt{arnaud96}). We fitted the spectra with a combination of different models to determine the shape of the direct and scattered broad-band continuum components (modeled with power laws), soft excess (modeled with a black body) and rest-frame line-of-sight X-ray absorption.  The models take into account the Galactic absorption using column densities taken from \cite{dickey1990}. We fitted both the pn and MOS spectra at the same time with the parameters of the model tied, except for the continuum normalization to take into account cross-calibration problems between MOS and pn cameras \citep{mateos09}.

To accept the detection of a model component, we used the F-test with a significance threshold of 95 per cent. The X-ray luminosities have been computed in the rest-frame 2-10 keV energy band. They are corrected for X-ray absorption both Galactic and intrinsic to the sources.

J00: It is X-ray unabsorbed. The 1$\sigma$ upper limit on the column density is 6.7$\times$10$^{20}$ cm$^{-2}$. The best-fitting model is a combination of a black body with temperature $kT$=0.16$^{+0.01}_{-0.02}$ keV and an unabsorbed power law with photon index \textit{$\Gamma$}=1.66$\pm$0.09. The X-ray spectra (MOS and pn) are shown in Fig. \ref{j0x} (left). For a sanity check, we computed the X-ray absorption with a fixed photon index of \textit{$\Gamma$}=1.9, the typical value for type-1 AGN (\citealt{caccianiga04}, \citealt{gaalbiati05}, \citealt{mateos05a}, \citealt{mateos05b}, \citealt{tozzi06}, \citealt{mateos10}, \citealt{corral11}). We still classify this object as a low absorption AGN (N$_{\rm H}\leq$1.3$\times$10$^{21}$ cm$^{-2}$). The black body emission is a phenomenological model often used in the literature \citep{corral15} to fit the soft X-ray excess. Nevertheless this is not physically correct for J00, as it is well known that the temperature of the black body for a SMBH such as the one for J00 (see Sec.~\ref{sec_bhmass}) is too low to explain the soft X-ray excess. We also tried a more physically motivated model, albeit more complex, replacing the black body model with a hot diffuse gas model (\texttt{mekal} model in XSPEC; \citealt{mewe85}, \citealt{mewe86}, \citealt{kaastra92}, \citealt{liedahl95}). This gives us an upper limit on the column density of 2.4$\times$10$^{20}$ cm$^{-2}$ (with \textit{$\Gamma$}=1.83$\pm$0.07, $\chi^2$/d.o.f=173.7/139). The results, in terms of the X-ray classification as absorbed/unabsorbed, do not change using one model or the other. Therefore we use in this paper the results with the black body since we obtain a more conservative value of N$_{\rm H}$.  The equivalent width of the Fe line at 6.4 keV in rest-frame is formally EW=0.30$^{+0.16}_{-0.17}$ keV, but adding this line is not statistically significant ($\Delta \chi^2$=3 for $\Delta$d.o.f.=1, a 1.73$\sigma$ detection). Looking at the spectrum of J00 (Fig.~\ref{j0x}) we can see a bump at the hard energies, but we believe that is not real, since it is only present in one of the EPIC cameras. This feature is probably associated with residuals in the background subtraction. Nevertheless, since the shape of the continuum is very well determined by the values at lower energies, this is not affecting our best-fit estimates.

J02: The best-fitting model is an absorbed power law with photon index \textit{$\Gamma$}=2.08$\pm$0.09 and intrinsic N$_{\rm H}$=1.7$_{-1.4}^{+2.0}\times$10$^{21}$ cm$^{-2}$. The EW of the Fe line is formally EW=1.12$^{+0.49}_{-0.39}$ keV, that is strong, but the detection is not significant ($\Delta \chi^2$=6 for $\Delta$d.o.f.=1, a 2.45$\sigma$ detection).

\section{UV-to-NIR Observations}
\label{sec_obs}

We have obtained UV-to-NIR high resolution spectra for both objects at the Very Large Telescope (VLT) with the VLT/XSHOOTER instrument \citep{vernet11}. The instrument divides the light in three paths that lead to three arms: one for the UV light, the second for the visible light and the last one for the near infrared (UVB, VIS and NIR, respectively). Each arm disperses the light with an echelle grating. The observations were taken with a 1.0$''\times$11$''$ slit for the UVB arm and 0.9$''\times$11$''$ slits for the VIS and NIR arms, respectively. Table~\ref{xsh_d} shows some technical details of the configuration setup of the observations.

\begin{figure}
 \includegraphics[width=8.5cm]{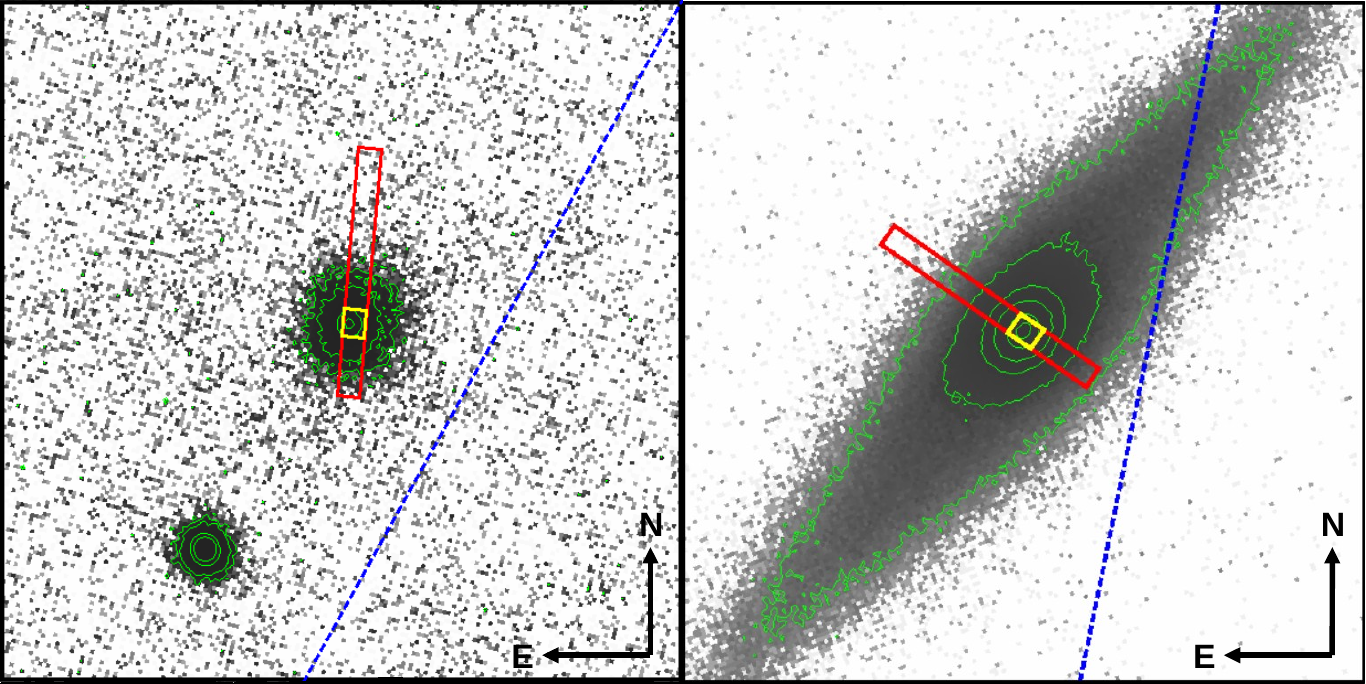}
    \caption{Section of 0.5$'\times$0.5$'$ of the acquisition image of J00 (left) and J02 (right) with the projection of the slit in the red box, 1$''\times$11$''$, for one of the nodding positions. Both images were taken with the i-band filter. The small yellow box inside the slit is the extraction region of the spectrum, 1$''\times$1.22$''$. The blue dashed line represents the parallactic angle. In green we plot contours of the objects. For J00 we plot 5, 10, 50 and 90 percent contour levels with respect to the peak of the AGN emission, and in J02 we plot the same contour levels plus an additional 2 percent one.}
       \label{adq}
 \end{figure}

The spectra were taken in nodding mode. In Table~\ref{xsh_d2} we show some information about the spectra acquisition. We present in Fig.~\ref{adq} the acquisition images of both objects and indicate the slit projected in the sky, the extraction region of the spectra and the parallactic angle. The observation dates were 2010-09-07 for J00 and 2010-09-04 for J02.

XSHOOTER is equipped with an Atmospheric Dispersion Correction (ADC) that allows the acquisition of the spectrum using any angle at any position in the sky. During our observations, the ADC was functional. For J02 we positioned the slit at an inclination angle close to the minor axis of the host galaxy to allow sky subtraction. Thanks to the ADC we could choose an inclination angle closer to the minor axis of the host galaxy. The observations were taken close to the meridian and with low air mass, meaning that the effect of the atmospheric dispersion is small.

The observations were reduced using the public XSHOOTER pipeline version 2.3.0 with Gasgano, following the instructions described in the XSHOOTER pipeline manual\footnote{ftp://ftp.eso.org/pub/dfs/pipelines/xshooter/xshoo-pipeline-manual-12.2.pdf}. We used a binning in the wavelength direction 0.02 nm/pix for the UVB and VIS arms and 0.06 nm/pix for the NIR arm, and in the slit direction 0.16 arcsec/pix for the UVB and VIS arms and 0.21 arcsec/pix for the NIR arm, as specified in the XSHOOTER user manual. We used the standard procedures in the pipeline. IRAF Laplacian Cosmic Ray Identification task for spectroscopy (\texttt{lacos\underline{ }spec}\textnormal, \citealt{vandokkum01}) was applied to the raw images for cosmic ray rejection and then each arm was reduced individually. We used the standard star GD71 (RA=05:52:27.61, DEC=+15:53:13.8) to calibrate the flux of our spectra. There are three different recipes to do the flux calibration: the offset mode, the staring mode and the nodding mode. We used the recipe for the staring mode, as it provides a better background subtraction.
We used the software IRAF to extract the 1D spectra being careful to follow the trace. This was carried out with the routine \texttt{apall}\textnormal. This is because the spectra have a trace whose centre was not constant in the cross-dispersion direction due to an imperfect order rectification of the pipeline.

\begin{table}
\centering
\caption{XSHOOTER observing configuration set up.}
\begin{tabular}{ | c || c | c | c | c | c | }
\hline
Arm &  Slit &  R &  Inst. Br. & T$_{exp}$  & Range \\
  (1) &  (2) &  (3) &  (4) & (5)  & (6) \\
\hline                     
UVB & 1.0 & 4350 & 0.86$\pm$0.20 & 1420 & 3000-5500 \\
VIS & 0.9 & 7450 & 1.14$\pm$0.25 & 1420 & 5500-10000 \\
NIR & 0.9 & 5300 & 0.88$\pm$0.08 & 2$\times$480 & 10000-25000 \\
\hline
\end{tabular}

$Notes$: (1): Instrument arm. (2): Slit width in arcsec. (3): Spectral resolution R=($\lambda / \delta \lambda$) according to the XSHOOTER webpage\footnote{https://www.eso.org/sci/facilities/paranal/instruments/xshooter/inst.html}. (4): Instrumental broadening in \AA, measured using arc lines. (5): Exposure time in s. (6): Wavelength coverage of each arm in \AA.
\label{xsh_d}
\end{table}

\begin{table}
\centering
\caption{XSHOOTER spectra acquisition information.}
\begin{tabular}{ | c || c | c | c |  }
\hline
Object &  Date & Nodding &  Airmass \\
  (1) &  (2) & (3) & (4)  \\
\hline                     
J00 & 2010-09-07  &  +2.5$''$ -2.0$''$  &   1.14 \\
J02 & 2010-09-04  & +2.5$''$ -1.5$''$  &   1.09 \\
\hline
\end{tabular}

$Notes$: (1): Object. (2): Observation date (YYYY-MM-DD). (3): Nodding separation in arc sec. (4): Airmass.
\label{xsh_d2}
\end{table}

We joined the spectrum from each XSHOOTER arm following the information about the dichroic crossover region in the XSHOOTER User Manual. To join the UVB and VIS arms, the crossover region is at 5595 \AA, while between the VIS and NIR it is at 10140 \AA. The transition regions are 5560-5638 \AA~and 10095-10350 \AA,~respectively. A 0.9-1.1 scaling factor between arms is sometimes needed to match in flux the complete spectrum (\citealt{lopez16}, \citealt{nisini16}). We used the continuum in these regions around the crossover points to compute the flux scaling factors for each arm using the VIS one as reference. We  scaled the UVB arm spectra using a factor of 0.9 while for the NIR arm a 1.0 flux scaling factor was acceptable. Errors were propagated through this process.

The aperture used to extract the spectra was defined to maximize the signal to noise of the AGN emission. This was carried out using the software IMFIT\footnote{http://www.mpe.mpg.de/$\sim$erwin/code/imfit/} on the acquisition images of the VLT observations. Both images were taken using the i-band filter. In both targets a bright nuclear source was detected (see contours in Fig.~\ref{adq}). To compute the fraction of AGN light that enters through the slit, for J00 we decomposed the emission in a Gaussian function for the AGN, and a Sersic profile for the host galaxy. J02 is an edge-on galaxy so we used a Gaussian profile for the AGN and two Sersic profiles for the host galaxy, to fit the bulge and the disk. The parameters of the Sersic and Gaussian models (intensity, $\sigma$, effective radius, Sersic index), are computed with errors of $\sim$20 percent or less, indicating reliable fits of the photometric images. In Fig. \ref{adq} the contours show that we have enough quality to fit the shape of the different components of our objects. We can trust the reliability of the acquisitions images to measure the slit losses using the width of the Gaussian profiles which are $\sim$2.5 and $\sim$3.5 pixels for J00 and J02, respectively.

We estimated that about 75 per cent and 56 per cent of the AGN emission are included in the slit for J00 and J02, respectively.

The reduction of the science spectra is concluded by correcting for Galactic extinction, putting the wavelength and flux in rest frame and converting the air wavelengths into vacuum wavelengths. Information to perform these corrections was obtained from the public data available in NASA/IPAC Extragalactic Database (NED).

\section{Analysis}
\label{sec_analysis}

In this section we describe the steps carried out to isolate the AGN emission in the XSHOOTER spectra and to determine the properties of the AGN and their host galaxies.

\subsection{AGN and host galaxy continuum decomposition}
\label{sec_decomp}

To decompose the extracted spectra into AGN and host galaxy emission at UV-to-NIR wavelengths we used the software STARLIGHT (\citealt{cid05}, \citealt{mateus06}).The best-fitting model was obtained by minimizing the $\chi^2$ statistic. We modeled the spectra with a composite model including a host galaxy spectrum plus an absorbed (by nuclear and host galaxy extinction) AGN spectrum. To reduce noise and to follow the recommendations of the STARLIGHT manual we rebinned the spectrum to 2 \AA~bins. SED decomposition is an approach commonly used to compute the relative contributions from the AGN and their hosts. Nevertheless, both the good spectral resolution and wide wavelength coverage of the X-SHOOTER spectra allow us to calculate the AGN and stellar components without resorting to a full SED decomposition.

We need to assume an spectral shape for the rest-frame UV to near-IR AGN continuum. There is substantial scatter in the continuum shapes of individual 
AGN which also depend on the SMBH mass and can vary with time (\citealt{koratkar99}, \citealt{schmidt12}, \citealt{baron16}). Since our aim is to reproduce the intrinsic AGN continuum by finding the model that best fits each source we have adopted the broken power law models (F$_{\lambda}\propto{\lambda}^{\alpha}$) from \cite{polletta07}. In these models the spectral index ranges from $\alpha$=-1.9 to -1.4 for $\lambda<$10000~\AA~and $\alpha$=-0.8 to -0.6 redwards, and the break is located at 10000~\AA. This break is likely associated with the change in the AGN continuum slope between the IR bump and the Big blue bump \citep{koratkar99}. We used this information to create a set of broken power laws with the previously mentioned index range to reproduce the intrinsic AGN continuum emission of our objects. The steps in the power law index to create the grid of broken power law models are $\Delta\alpha$=0.125 for the blue region and $\Delta\alpha$=0.05 for the red.

The next parameter needed in our fit is the obscuration of the nuclear region of the AGN. We used the extinction model of the Small Magellanic Cloud (SMC; \citealt{gordon03}) as it is the one that fits better the AGN spectra \citep{hopkins04}. We tried different extinction models but the SMC is the one that minimized the $\chi^2$. In particular we checked the Large Magellanic Cloud Super-Shell and Average models from \cite{gordon03}, the \cite{calzetti00} law, the Milky Way model form \cite{allen76}, and the model from \cite{cardelli89}. In addition, using the SMC model we obtained the most conservative A$_{\rm V}$ values. We constrained the nuclear extinction to be between A$_{\rm V}$=10 mag and A$_{\rm V}$=0 mag.

The host galaxy contribution is modeled using the Single Stellar Population (SSP) templates from the \cite{bruzual03} library. We used 45 models with metallicities $Z$=0.05, 0.02 and 0.004 in units of the solar value and ages ranging from 1 Myr to 13 Gy. We obtained SSP in the 0.8-13 Gyr range for our objects.

\begin{figure*}
 \centering
 \includegraphics[width=0.9\textwidth ]{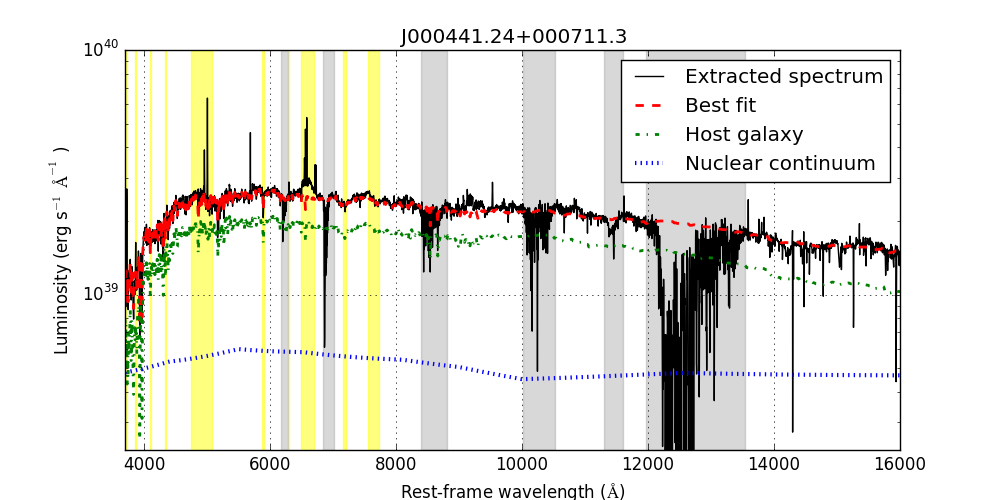}
 \includegraphics[width=0.9\textwidth ]{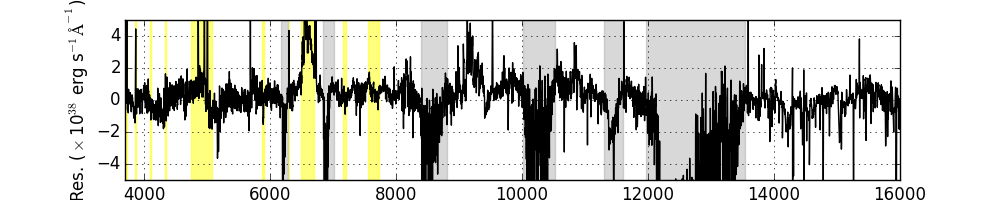}
 \includegraphics[width=0.9\textwidth ]{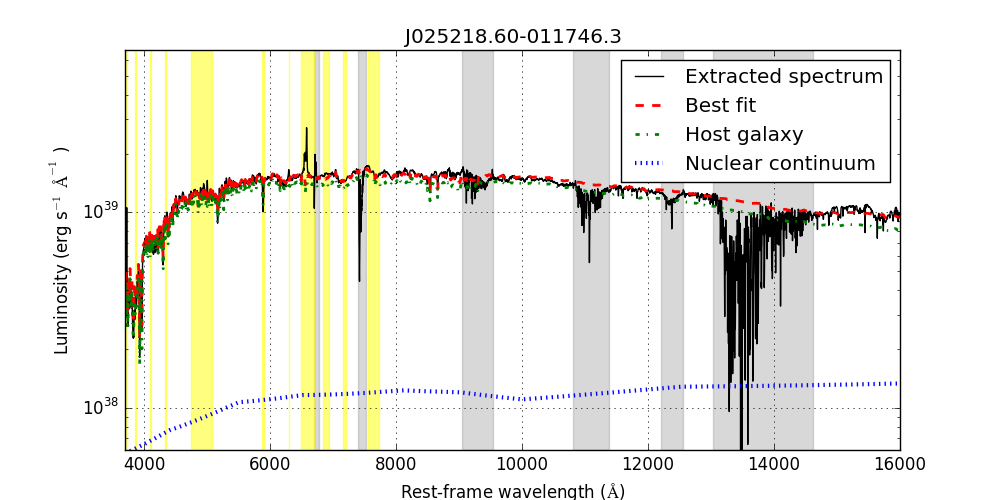}
 \includegraphics[width=0.9\textwidth ]{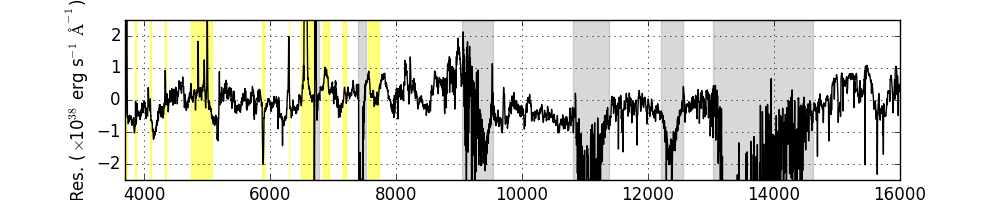}
    \caption{Decomposition of the UV-to-NIR spectra of J00 (top) and J02 (bottom) into host galaxy and AGN components. The upper panels represent the results of the fits while the bottom panels show the residuals. The grey and yellow bands indicate the telluric absorption features and the spectral regions with AGN emission lines, respectively, which are ignored in the fits.  }
       \label{mo}
 \end{figure*}

In order to fit the AGN and host galaxy continuum emission only, we masked out the spectral ranges where the AGN emission lines and the telluric absorption lines are expected (see grey and yellow bands in Fig.~\ref{mo}). We excluded the regions between 6850-6950 \AA, 7165-7210 \AA~and 7550-7725 \AA,  where some residuals of telluric features in the Bruzual \& Charlot library are present.

The final spectral range used in the fit is between 3700 and 16000 \AA~(rest-frame).

Fig.~\ref{mo} shows the  results of the spectral decomposition. We see that in both objects, the total emission is dominated by the host galaxy. The preferred models are $\alpha_{\rm blue}$=-1.90 and $\alpha_{\rm red}$=-0.6 for J00 and for J02 $\alpha_{\rm blue}$=-1.78 and $\alpha_{\rm red}$=-0.6. The resulting values of intrinsic absorption associated to the AGN emission are A$_{\rm V}$=2.04$\pm$0.30 mag and  A$_{\rm V}$=2.19$\pm$0.33 mag for J00 and for J02, respectively.

We computed the errors in the extinction by varying the power-law index and then fitting again and calculating the $\chi^2$ statistics. The errors obtained are very small. We added an additional 15 percent error contribution to account for the overall uncertainty of the SMC model used \citep{gordon03}. This contribution to the error is the one that dominates.

We believe that the results from the AGN and host galaxy decomposition reported in this section are robust. The XSHOOTER spectra have a sufficiently large wavelength range to constrain in a robust way the host galaxy contribution, which is the major contribution at optical wavelengths. It is clear that in the spectra there are several stellar features that help constraining the emission from the hosts (see Fig.~\ref{mo}). If more relative contribution of the nuclei is present in the spectra, this would in fact, flatten the stellar features to a more featureless contribution. The STARLIGHT software fits this stellar features and the results in this paper are the best fitting ones.

\subsection{Narrow line and broad line Balmer decrements}
\label{sec_bdec}

After removing the host galaxy component and after having taken into account the slit losses, we estimated the Balmer decrement by using the broad and narrow components of the H$_{\alpha}$ and H$_{\beta}$ emission lines.

We fitted the nuclear spectra from rest-frame 6270 to 6800 \AA~for H$_{\alpha}$. For the H$_{\beta}$ region we fitted from rest-frame 4600 to 5050 \AA~for J00 and from 4800 to 5050 \AA~for J02. We used a wider wavelength range in the first object to ensure fitting the whole broad H$_{\beta}$ emission line. To model the AGN narrow emission lines we used Gaussian functions, assuming that they share the same width in velocity space. Hydrogen broad emission lines are fitted with Gaussian profiles as well. We included also a parameter for the broad line that is a shift of the centre of the Gaussian with respect to the vacuum values. This is because some AGN show broad lines with an offset, sometimes of a thousand of km/s due to strong winds (\citealt{sulentic00}, \citealt{steinhardt12}, \citealt{gaskell13}). For J00 the shift is $\sim$30 \AA~($\sim$1350km/s in velocity) while for J02 is $\sim$3 \AA~($\sim$150km/s), not unusual with respect to observed shifts \citep{sulentic00}. We used a power law to fit the continuum around the lines. The line parameters were obtained with the CIAO's SHERPA fitting tool \citep{freeman01}.

Since the broad H$_{\beta}$ component was not detected in any of the spectra, only an upper limit could be computed for the Balmer decrement from the BLR. Line parameters and Balmer decrements are reported in Table~\ref{halfa}. The narrow line Balmer decrement will be compared in Sec. \ref{sec_gastodust} with the absorption in the X-rays and the extinction of the UV-to-NIR continuum.

\begin{figure*}
 \centering
 \includegraphics[width=0.45\textwidth ]{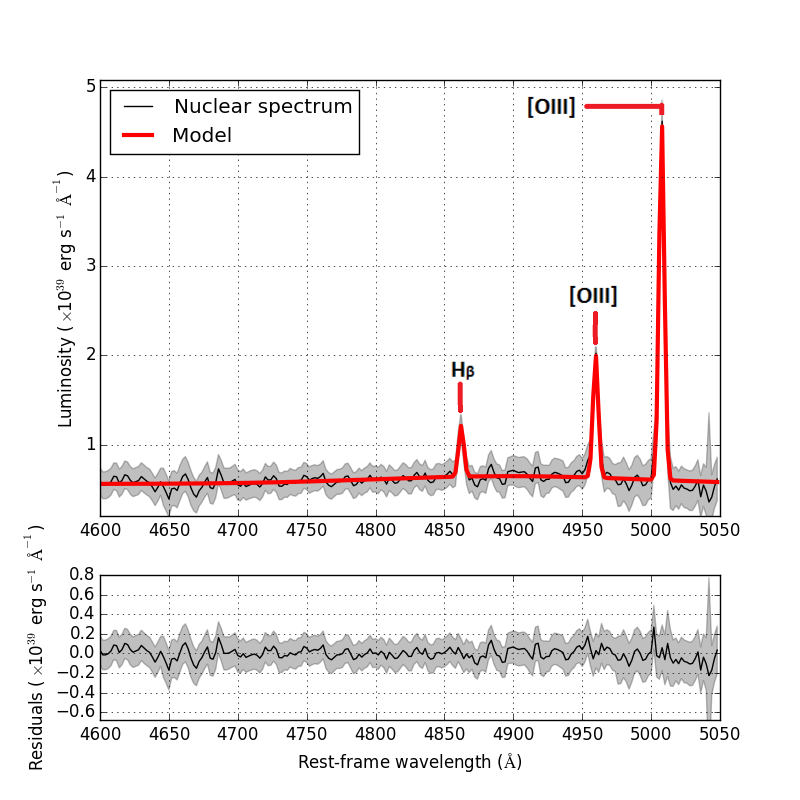}{\vspace{0cm}}
 \includegraphics[width=0.45\textwidth ]{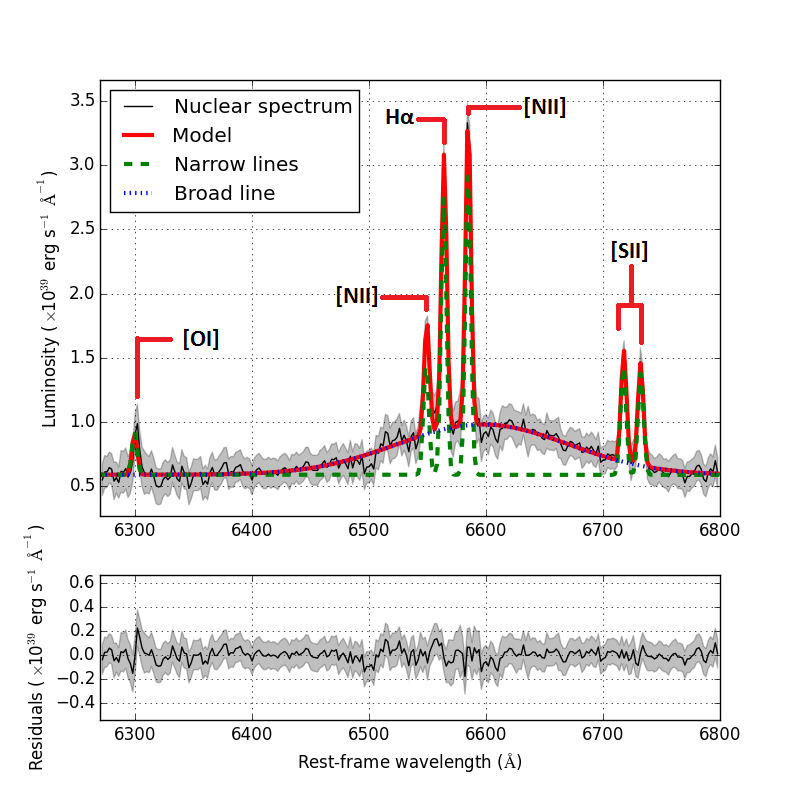}{\vspace{0cm}}
 \includegraphics[width=0.45\textwidth ]{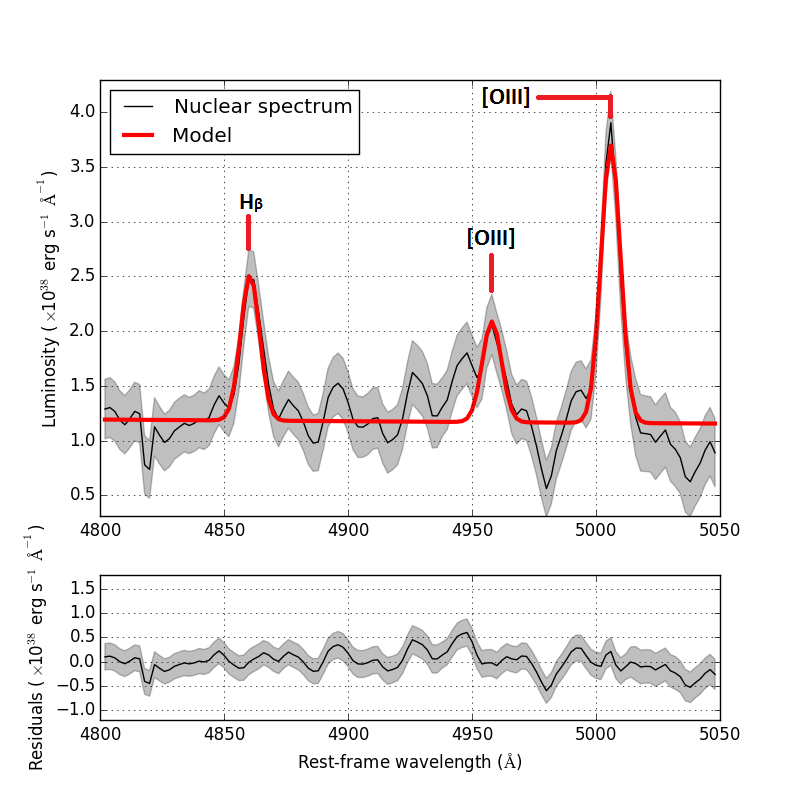}{\vspace{0cm}}
 \includegraphics[width=0.45\textwidth ]{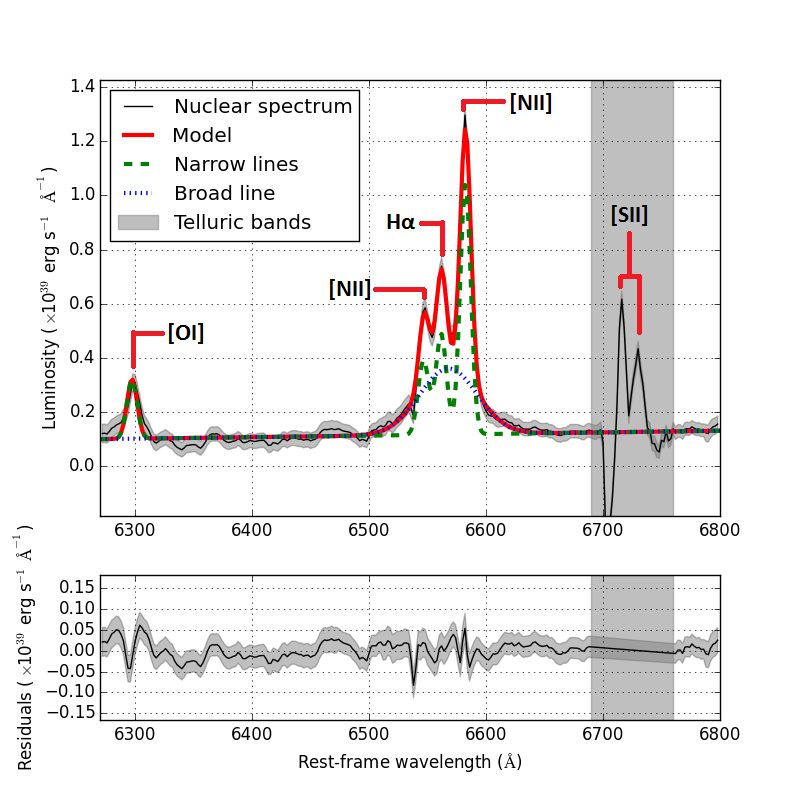}{\vspace{0cm}}
    \caption{Top: Decomposition of the AGN emission into continuum plus narrow and broad emission lines for J00 (top) and J02 (bottom). Left panels: H$_{\beta}$+[OIII] region. Right panels: H$_{\alpha}$ region. In grey we plot the error of each spectrum. We also indicate the observed narrow emission lines.}
       \label{lin}
 \end{figure*}

In Fig. \ref{lin} we show our results while best-fitting values are indicated in Table \ref{halfa}.

\subsection{SMBH masses}
\label{sec_bhmass}

From the XSHOOTER nuclear spectra, we derived an estimate of the SMBH masses for our targets. There are many different ways to compute SMBH masses \citep{trippe15}. We used the luminosity and FWHM of the H$_\alpha$ broad emission line and the expression from \cite{greene05},

\begin{equation}
\begin{split}
log \left( \frac{M_{SMBH}}{M_{\odot}} \right)  = (0.45 \pm 0.05)  \log \left( \frac{L_{H \alpha}}{10^{42}ergs~ s^{-1}}\right)+\\
+ (2.06 \pm 0.06) log \left(\frac{FWHM_{H \alpha}}{10^3 km~ s^{-1}} \right) + 6.40^{+0.09}_{-0.07}
\end{split}
\label{smbh}
\end{equation}
where M$_{\rm SMBH}$/M$_{\odot}$ is the mass of the SMBH in solar units. L$_{H \alpha}$ is the intrinsic luminosity of the H$_{\alpha}$ emission line in erg~s$^{-1}$, this is, corrected for both the nuclear and the host galaxy extinction, computed from the spectral fits. Finally FWHM$_{H \alpha}$ is the FWHM of the broad H$_{\alpha}$ emission line in km~s$^{-1}$. We also have corrected the FWHM of the lines by the instrumental spectral dispersion by subtracting the instrumental broadening in quadrature.

\begin{table}
\centering
\caption{AGN fitting results.}
\begin{tabular}{ | c | c | c |  }
\hline
$ $ & J00 & J02 \\
\hline
log(M$_{\rm SMBH}$/M$_{\odot})$ &  7.96$^{+0.23}_{-0.25}$ & 6.74$^{+0.26}_{-0.22}$ \\
FWHM$_{H\alpha \rm ,B}$ & 7830 $\pm$ 1221  & 2499 $\pm$ 175  \\
FWHM$_{H\alpha \rm ,N}$ & 246 $\pm$ 19  & 376 $\pm$ 17  \\
L$_{H\alpha \rm ,B}$  & 22.0$\pm$2.0 & 6.15$\pm$0.61 \\
L$_{H\alpha \rm ,N}$  &  2.63$\pm$0.13 & 0.52$\pm$0.08 \\
L$_{H\beta \rm ,B}$  & $\leq$16.2 & $\leq$1.51\\ 
L$_{H\beta \rm ,N}$  & 0.89$^{+0.19}_{-0.21}$ & 0.17$\pm$0.03 \\
L$_{\rm [OIII],5008\AA}$  & 8.37$^{+0.49}_{-0.62}$ & 0.81$^{+0.8}_{-0.12}$ \\  
L$_{H\alpha \rm ,N}^o$/L$_{H\beta \rm ,N}^o$ & 4.41$_{-1.34}^{+0.92}$ & 3.34$_{-1.05}^{+0.80}$  \\
L$_{H\alpha \rm ,B}^o$/L$_{H\beta \rm ,B}^o$ & $\geq$3.13 & $\geq$11.40 \\
 L$_{\rm bol}$ & 1192$\pm$ 118 &  243 $\pm$ 30 \\
 L$_{\rm Edd.}$ & 1.19$^{+0.97}_{-0.56}\times$10$^{6}$   &  0.071$^{+0.052}_{-0.036}\times$10$^{6}$  \\
L$_{\rm bol}$/L$_{\rm Edd.}$ &  0.0010$^{+0.0004}_{-0.0010}$  &  0.0034$^{+0.0011}_{-0.0034}$ \\
\hline

\end{tabular}

$Notes$: SMBH masses (in log(M$_{\rm SMBH}$/M$_{\odot}$) units) and properties of the different broad and narrow emission lines used in our analysis. The full width at half-maximum (FWHM) is in km/s. All luminosities are in units of 10$^{40}$ erg/s. The Balmer decrement, as indicated in the text, is calculated using the reddened AGN spectrum. All luminosities are extinction corrected, except those with the superindex 'o'.
\label{halfa}
\end{table}

We show the SMBH masses computed using Eq.~\ref{smbh}, as well as the relevant luminosities, in Table \ref{halfa}. In addition we show the bolometric and Eddington luminosities for our sources, as well as the Eddington ratio (L$_{bol}$/L$_{Edd.}$). The bolometric luminosities are calculated using the expression L$_{\rm bol}$=9$\times \lambda$L$_{\rm 5100}$ \citep{kaspi00}, being L$_{\rm 5100}$ the monochromatic luminosity of the unreddened nuclear emission at rest-frame 5100 \AA. The Eddington ratio is within the expected values for nearby AGN with similar bolometric luminosities and SMBH masses \citep{panessa06}.

\subsection{Host galaxy masses}
\label{sec_hostmass}

It is well known that the SMBH mass and the spheroidal mass of the host galaxies follow a linear relation (\citealt{merrit01}, \citealt{park12}). In this study we calculate both the dynamical mass and the stellar mass from the spheroidal component of the host galaxy.

\subsubsection{Stellar masses}
\label{sec_mstell}

STARLIGHT provides stellar masses for its best fit models. We have corrected those for slit losses using the spheroidal components obtained by IMFIT (see Sec.~\ref{sec_obs}): the fractions of the spheroid light that went through the slit are 9.0 and 6.2 percent for J00 and J02, respectively.

The stellar mass derived using STARLIGHT is computed using mass-to-light relations, converting each SSP contribution to stellar mass. STARLIGHT does not compute errors in the best-fitting values. In \cite{bell01} it is discussed the stellar mass-to-light ratio and the uncertainties associated to this ratio. This uncertainties are all of order 0.1-0.2 dex. We use the largest value as the error for our values to be conservative. Our results are shown in Table \ref{masses}.

 \begin{figure*}
 \centering
 \includegraphics[width=8.5cm]{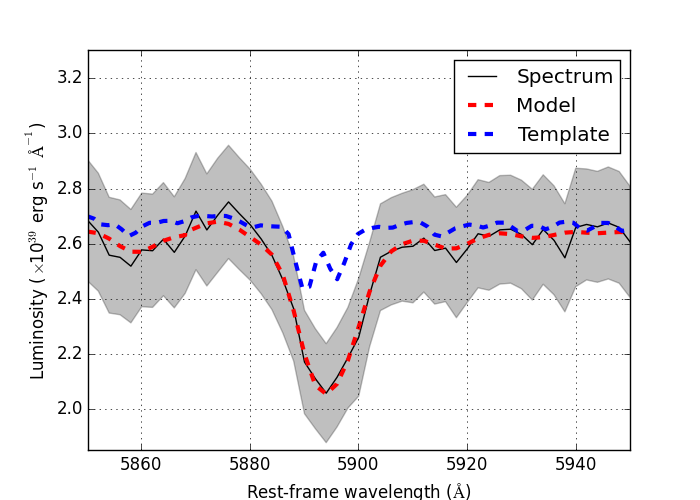}
 \includegraphics[width=8.5cm]{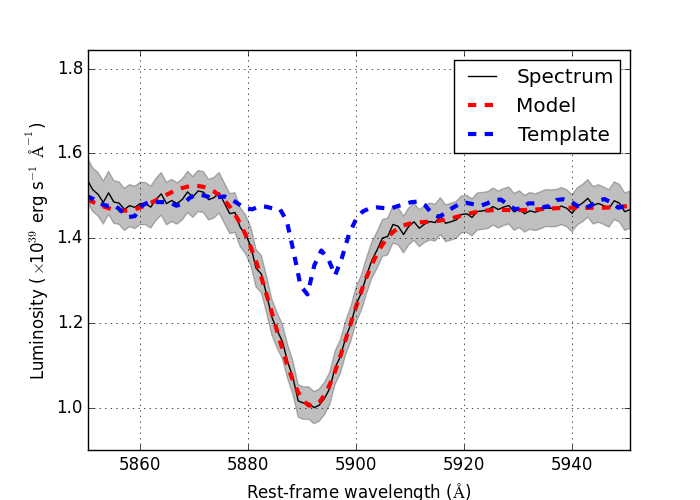}
    \caption{Spectrum (in black) of the Na I Doublet of J00 (left) and J02 (right) and its fit (in red). In grey we show the error of each spectrum. In blue we plot the template used to measure $\sigma_{e}$ in the region of NaID. The template has a spectral resolution of 1.28\AA.}
       \label{na}
 \end{figure*}
\subsubsection{Dynamical masses}
\label{sec_mdyn}

As a sanity check, we also computed the dynamical mass of the spheroidal component of the host galaxies. We used the relation between the line of sight velocity dispersion ($\sigma_{e}$) and the dynamical mass.

The observed width of the NaI doublet at rest-frame $\lambda\lambda$5896, 5890 \AA~(Na\,{\sc I}D) is a stellar absorption feature that can be used to calculate $\sigma_{e}$ \citep{spiniello12}. The NaID feature is a convolution of the resolution of the instrument, the intrinsic width of the stellar population and $\sigma_{e}$. To compute $\sigma_{e}$, we used the instrumental resolution from Table~\ref{xsh_d}. For the stellar population, we used the Single Stellar Population (SSP) template from the Bruzual and Charlot 2003 library \citep{bruzual03} that based on the results of STARLIGHT, is the most dominant SSP for each host galaxy: the SSP of 0.9 Gyr and $Z$=0.05 for J00, and 11 Gyr and $Z$=0.02 for J02.  The standard dispersion of the Gaussian function gives the desired $\sigma_{e}$.

In Fig. \ref{na} we show the NaID and the results of the fits. We plot in this figure the template used in each object with the spectral resolution at the region of the doublet. The spectral resolution is computed using the FWHM of the closest arc line of the XSHOOTER observation. To calculate the dynamical mass of our AGN hosts we used the Virial relations from \cite{cappellari06} and \cite{taylor10}.

\begin{equation}\label{mdynf}
M_{dyn}=k(n) \times \frac{r_{e} \sigma_{e}^{2}}{G}
\end{equation}
where $r_{e}$ is the effective radius of the spheroidal mass of the host galaxy, $\sigma_{e}$ is the line-of-sight velocity dispersion, the factor $k(n)$ that depends on the Sersic index and $G$ is the universal gravitational constant. Using the software IMFIT we obtained $r_{e}$ and the Sersic index by fitting the acquisition images from the VLT. To compute the error in the spheroidal mass we propagate errors.

We summarize in Table \ref{masses} the properties of our AGN hosts. We clearly see that our dynamical mass estimates are compatible with the stellar masses calculated before. The studied galaxies are massive but not atypical \citep{vitale13}.

\begin{table}
\centering
\caption{Host galaxy properties.}
\begin{tabular}{ | c | c | c |  }
\hline
 $ $ & J00 & J02 \\
\hline
 $\sigma_{e}$  & 145$\pm$109 km/s & 260$\pm$87 km/s \\
 $r_{e}$ & 4.21$\pm$0.39 kpc & 0.84$\pm$0.04 kpc \\
 $n$ & 1.17 & 0.96 \\ 
 $k(n)$ & 7.92 & 7.96 \\ 
 log(M$_{dyn}$/M$_{\odot})$ & 11.07$^{+0.23}_{-0.42}$ & 10.85$^{+0.11}_{-0.12}$ \\
 log(M$_{stell}$/M$_{\odot})$ & 10.73$^{+0.20}_{-0.20}$ & 10.89$^{+0.20}_{-0.20}$ \\
\hline
\end{tabular}

$Notes$: Line-of-sight velocity dispersion ($\sigma_{e}$) of the two AGN. The value $r_{e}$ is the effective radius. The factor $k(n)$ in Eq.~\ref{mdynf} depends on $n$, that is the Sersic index. The spheroidal masses in logarithmic units of M$_{\odot}$.
\label{masses}
\end{table}

\section{Discussion}
\label{sec_discuss}

In the following subsections we discuss one by one the possible causes of the apparent discordant properties of our AGN in the UV/optical range and in X-rays.

\subsection{Compton-thick or Compton-thin obscuration}
\label{sec_ct}

Based on the observed X-ray properties, we can rule out that our objects are Compton-thick. First of all the broad band continuum shape is too steep to be produced mainly by reflected emission as demonstrated in Sec.~\ref{sec_x}. Compton-thick sources usually have $\Gamma\sim$1.0 (\citealt{brightman99}, \citealt{winter08}, \citealt{georgantopoulos11}, \citealt{delmoro16}).

Compton-thick AGN are expected and known to display large EW Fe K emission lines at 6.4 keV ($\leq 1$~keV; \citealt{gandhi14}), due to the highly suppressed underlying continuum. No Fe line is detected by the fits with high significance in any of our sources.

The Compton-thin nature of the sources is supported by the high L$_{2-10~keV}$/L$_{[\rm O\,{\sc III}]}$ ratio, where L$_{[\rm O\,{\sc III}]}$ is the unreddened luminosity of the [O\,{\sc III}] emission line at rest-frame $\lambda$5007 \AA. This is because if we only detect in X-rays the soft scattered component (which is only a few percent of the intrinsic AGN power), the X-ray luminosity can be largely underestimated. Therefore we can use L$_{[\rm O\,{\sc III}]}$ as a proxy of the bolometric luminosity, and compare it to the L$_{2-10~keV}$.  Compton-thick sources have L$_{2-10~keV}$/L$_{[\rm O\,{\sc III}]}<$0.1-1.0~(\citealt{bassani99}, \citealt{akylas09}). This ratio is 69 for J02 and 22 for J00. This effectively excludes the Compton-thick character of both of our sources.

Alternatively, we could use the mid-IR to L$_{X}$ ratio to identify Compton-thick obscuration (\citealt{gandhi09}, \citealt{asmus11}, \citealt{mateos15}, \citealt{stern15}). If we use the three shortest $\lambda$-bands of the Wide-Field Infrared Survey Explorer (WISE, 3.4, 4.6, 6, 12 $\mu$m; \citealt{wright10}), we find that our sources have mid-IR colours that fall outside of the region occupied by AGN \citep{mateos12}, because their catalogued fluxes have significant contamination from the AGN hosts.

\subsection{Host-SMBH relations}
\label{sec_msigma}

We estimated the SMBH-to-host galaxy mass ratio for our two sources (see Table~\ref{res_t}) and we compared them with the value reported by \cite{merrit01}, $\langle \log(M_{\rm SMBH}/M_{bulge}) \rangle$=-2.9  with a rms of 0.5 dex. This has been derived by using the samples of bulges and elliptical galaxies of \cite{ferrarese00} and \cite{gebhardt00}. For  J00 we obtain $\langle \log(M_{\rm SMBH}/M_{bulge}) \rangle$=-2.77, which is consistent with the Meritt \& Ferrarese relation. This means that the galaxy is as massive as expected by its SMBH. J02 has $\langle \log(M_{\rm SMBH}/M_{bulge})\rangle$=-4.15, that is more than 2 times the rms below the standard relation. For J02 host galaxy dilution could explain, at least in part, the lack of broad emission line detection in the 6dF spectrum. This is because we expect more impact of the star-light dilution on the AGN emission compared to AGN with less massive host galaxies. There are many examples in the literature of apparently normal galaxies (eg. XBONGs) in which, after the host galaxy contamination is removed, AGN emission is revealed (\citealt{severgnini03}, \citealt{georgantopoulos05}, \citealt{caccianiga07}). In the next sections we will investigate if this is the only factor that contributes to the observed optical/X-ray discrepancy.

\begin{table}
\centering
\caption{Gast-to-dust and M$_{\rm SMBH}$/M$_{Host}$ ratios.}
\begin{tabular}{ | c || c | c | }
\hline
Object &  A$_{\rm V}$/N$_{\rm H}$  & $\log(M_{\rm SMBH}/M_{bulge})$  \\
\hline          
J00 &  $\geq$2.61$\times$10$^{-21}$  &  -2.77$^{+0.36}_{-0.34}$  \\
J02 &  1.30$_{-1.1}^{+1.8}\times$10$^{-21}$  &  -4.15$\pm$0.39  \\
Reference value &  5.3$\times$10$^{-22}$ &  -2.9$\pm$0.5  \\
\hline
\end{tabular}

$Notes$: The units of A$_{\rm V}$/N$_{\rm H}$ are in mag cm$^{-2}$. The reference value for A$_{\rm V}$/N$_{\rm H}$ is the Galactic and for $\log(M_{\rm SMBH}/M_{bulge})$ is the mean value from \cite{merrit01}.
\label{res_t}
\end{table}

\subsection{Dust-to-gas ratio of the obscuring medium and Balmer decrement}
\label{sec_gastodust}

Typically AGN have dust-to-gas ratios lower than or compatible with the Galactic value (\citealt{maiolino01}, \citealt{vasudevan09}, \citealt{parisi11}, \citealt{marchese12}, \citealt{hao13}, \citealt{burtscher16}). This result does not appear only in X-ray selected studies, but also on optical and IR selected AGN (\citealt{wilkes02}, \citealt{young08}). We have compared A$_{\rm V}$ (see Table~\ref{nh}) and N$_{\rm H}$ (see Table~\ref{data}) for our sources. For J00 A$_{\rm V}$/N$_{\rm H}\geq$2.61$\times$10$^{-21}$  mag cm$^{-2}$, while for J02 A$_{\rm V}$/N$_{\rm H}$=1.30$_{-1.1}^{+1.8}\times$10$^{-21}$ mag cm$^{-2}$. The Galactic relation is A$_{\rm V}$/N$_{\rm H}$=5.3$\times$10$^{-22}$ mag cm$^{-2}$. J00 shows an A$_{\rm V}$/N$_{\rm H}$ more than 5 times the Galactic value. For J02 the value is consistent with the Galactic.
As mentioned in Sec.~\ref{sec_decomp}, the SMC extinction model is one of the most conservative measurements on the A$_{\rm V}$ of all models taken into account. The results provided by the other extinction models do not change the results in terms of the dust-to-gas ratio.

A dust-to-gas ratio higher than the Galactic value explains the observed properties of J00. It is a scenario offered in \cite{panessa02}, to explain the different optical and X-ray classification of unabsorbed Seyfert 2. In this case a strong contribution of dust has less effect in the X-ray emission than the optical one. There are a few examples in the literature of higher dust-to-gas than the Galactic value (\citealt{trippe10}, \citealt{mehdipour12}, \citealt{huang12}), but this scenario is not very common as it is about 3-9 per cent of the sources (\citealt{maiolino01}, \citealt{caccianiga04}, \citealt{malizia12}).


\begin{table}
\centering
\caption{AGN optical extinction.}
\begin{tabular}{ | c || c | c | }
\hline
$ $ &  J00 & J02  \\
\hline          
A$_{\rm V,NLR}$ & 1.07$^{+0.67}_{-0.81}$ & 0.30$^{+0.77}_{-0.81}$   \\
A$_{\rm V,cont.}$ & 2.04$\pm$0.30 & 2.19$\pm$0.33  \\
A$_{\rm V,X-ray}$ & $\leq$0.05 & 0.13  \\
\hline
\end{tabular}

$Notes$: Optical extinction of the NLR and AGN continuum in mag. The NLR extinction is converted through the Balmer decrement (\citealt{bassani99}, \citealt{pappa01}, \citealt{carrera04}) through the expression E(B-V)=2.07$\times$log((H$_{\alpha}$/{H$_\beta$})/3) and R$_{\rm V}$=3.1. The A$_{\rm V,X-ray}$ is the optical extinction corresponding to the N$_{\rm H}$ column density using the SMC model of \citet{gordon03}.
\label{nh}
\end{table}

\subsection{Intrinsically weak BLR emission}

To investigate whether our AGN have a BLR with non-standard properties (e.g. underluminous) we have determined the luminosity ratio L$_{\rm NLR}$/L$_{\rm BLR}$ from the broad and narrow components of the H$_\alpha$ emission line (see Table~\ref{halfa}). Then we compare our values with the relation between L$_{\rm NLR}$/L$_{\rm BLR}$ and L$_{\rm BLR}$ found for AGN of similar z and luminosities to ours from \cite{stern12}. Our objects have L$_{\rm NLR}$/L$_{\rm BLR}$ ratios within the observed 0.4 dex scatter hence, none of our sources appears to have intrinsically weak BLR.

\subsection{Variability}
\label{sec_var}
AGN are highly variable sources across the full electromagnetic spectrum at both long and short time scales (\citealt{ulrich97}, \citealt{mateos07}, \citealt{krumpe10}, \citealt{garcia15}, \citealt{hernandez15}, \citealt{lamassa15}). This is originated by variability in the accretion rate and by extinction variability in the line-of-sight material (\citealt{markowitz14}, \citealt{miniutti14}). As the X-ray and UV/optical observations have not been taken simultaneously we cannot rule out that this might also contribute to the observed mismatch between the optical and X-ray properties of our AGN.

As a check, we carried out the UV-to-optical spectral decomposition into AGN and host galaxy emission with the public SDSS-DR7 spectrum of J00 (taken in 2000-09-05), using the same broken power law for the AGN emission and the same host galaxy emission from Sec.~\ref{sec_decomp}. In this test we compare the intrinsic flux of the AGN, so we have taken into account that the SDSS spectra were taken with a 3$''$ fiber and hence a higher fraction of AGN enters through the fiber and the host galaxy contribution is higher than in the XSHOOTER spectra. Analyzing the results there is some variation, that is best fitted by an extinction change (A$_{\rm V,SDSS}$=0.69$\pm$0.10 mag versus the one obtained by XSHOOTER, that is A$_{\rm V,XSH}$=2.04$\pm$0.29 mag) instead of a change in the emission of the AGN.  The computed A$_{\rm V}$ gives a higher dust-to-gas ratio than the Galactic (A$_{\rm V}$/N$_{\rm H}>$8.8$\times$10$^{-22}$ mag cm$^{-2}$), as with the XSHOOTER data. The 6dF optical spectrum of J02 is not of sufficient quality for conducting a spectral decomposition analysis.

We conclude that extinction variability could be present in at least one of the sources (the one with the largest A$_{\rm V}$/N$_{\rm H}$ ratio). Even so, a higher than Galactic dust-to-gas ratio is also needed in that source. For this source we only observe a marginal variation of the intrinsic flux of 1.6$\sigma$. Simultaneous X-ray and optical observations are needed to assess the actual importance of variability.

\section{Conclusions}
\label{sec_conclusions}
In this work we have investigated the origin of the apparent mismatch of the optical and X-ray classifications of two AGN with high optical extinction but low X-ray absorption. In our two selected objects there is a clear broad line in H$_{\alpha}$ using the data from VLT/XSHOOTER after a careful removal of the host galaxy contribution.

We discussed several scenarios that could explain the discordance of our observations. We ruled out a Compton-thick nature of our sources, on the basis of the L$_{2-10~keV}$/L$_{\rm [OIII]}$ ratio. We also discarded that the total or partial absence of broad lines in the spectrum is caused by an intrinsically weak BLR emission.

The origin of the mismatch for each object is found to be different. An intrinsically different A$_{\rm V}$/N$_{\rm H}$ is the best explanation for the properties of the object J00 without the need to invoke variability. This obscuring material has a dust-to-gas ratio really different than the majority of the AGN population. The other object, J02, has a massive host galaxy in comparison with its SMBH, so that the broad emission lines and the nuclear continuum are swamped by the host galaxy star-light which makes them very difficult to detect.

\section*{Acknowledgments}
We thank the anonymous referee for the comments, which have helped us to improve the quality of the manuscript. We thank R. Antonucci and J. Stern for their references and comments that improved our manuscript. IO-P thanks A. Khan-Ali for his support with SHERPA and \texttt{PYTHON}.

IO-P and FJC acknowledge financial support through grant AYA2015-64346-C2-1-P (MINECO/FEDER).
SM acknowledges financial support by the Spanish Ministry of Economy and Competitiveness through grant AYA2016-76730-P (MINECO/FEDER).
AC, RD, PS and VB acknowledge financial support by the Italian Space Agency (contract ASI-INAF
I/037/12/0).

Based on observations obtained with XMM-Newton, an ESA science mission with instruments and contributions directly funded by ESA Member States and NASA.

Based on observations collected at the European Organisation for Astronomical Research in the Southern Hemisphere under ESO programme 085.B-0757(A).

This research has made use of the NASA/IPAC Extragalactic Database (NED) which is operated by JPL, Caltech, under contract with the National Aeronautics and Space Administration. The CASSIS is a product of the Infrared Science Center at Cornell University, supported by NASA and JPL.

Funding for the SDSS and SDSS-II has been provided by the Alfred P. Sloan Foundation, the Participating Institutions, the National Science Foundation, the U.S. Department of Energy, the National Aeronautics and Space Administration, the Japanese Monbukagakusho, the Max Planck Society, and the Higher Education Funding Council for England. The SDSS Web Site is http://www.sdss.org/.

The SDSS is managed by the Astrophysical Research Consortium for the Participating Institutions. The Participating Institutions are the American Museum of Natural History, Astrophysical Institute Potsdam, University of Basel, University of Cambridge, Case Western Reserve University, University of Chicago, Drexel University, Fermilab, the Institute for Advanced Study, the Japan Participation Group, Johns Hopkins University, the Joint Institute for Nuclear Astrophysics, the Kavli Institute for Particle Astrophysics and Cosmology, the Korean Scientist Group, the Chinese Academy of Sciences (LAMOST), Los Alamos National Laboratory, the Max-Planck-Institute for Astronomy (MPIA), the Max-Planck-Institute for Astrophysics (MPA), New Mexico State University, Ohio State University, University of Pittsburgh, University of Portsmouth, Princeton University, the United States Naval Observatory, and the University of Washington.

IRAF is distributed by the National Optical Astronomy Observatory, which is operated by the Association of Universities for Research in Astronomy (AURA) under cooperative agreement with the National Science Foundation.

The STARLIGHT project is supported by the Brazilian agencies CNPq, CAPES and FAPESP and by the France-Brazil CAPES/Cofecub program.

This publication makes use of data products from the Two Micron All Sky Survey, which is a joint project of the University of Massachusetts and the Infrared Processing and Analysis Center/California Institute of Technology, funded by the National Aeronautics and Space Administration and the National Science Foundation.

\label{lastpage}
\end{document}